\documentclass[preprint]{iucr}

\usepackage{color,xcolor}
\usepackage{xspace}
\RequirePackage{graphicx}
\usepackage{epstopdf}
\usepackage{mathtools, amsmath, amsthm, amssymb, amsfonts,bm}
\usepackage[12pt]{moresize}
\usepackage{multirow}
\usepackage{enumitem}
\usepackage{url}

\usepackage{threeparttable}
\usepackage{mathtools,xparse,amsmath}

\graphicspath{{./figures/}}

\definecolor{tagcolor}{HTML}{21a889}

\definecolor{dgreen}{HTML}{008000}

\newcommand{\be}{\begin{enumerate}[wide, labelwidth=!, labelindent=0pt,
        label=\textbf{\textcolor{blue}{\arabic*}.}]}
    \newcommand{\bei}{\begin{enumerate}}
        \newcommand{\ee}{\end{enumerate}}
    \newcounter{saveenumi}

\newcommand{\dd}{\mathrm{d}}

\newcommand{\nba}[1]{}

\newcommand{\qmax}{\ensuremath{Q_{\mathrm{max}}}\xspace}
\newcommand{\qmin}{\ensuremath{Q_{\mathrm{min}}}\xspace}
\newcommand{\rmin}{\ensuremath{r_{\mathrm{min}}}\xspace}
\newcommand{\rmax}{\ensuremath{r_{\mathrm{max}}}\xspace}

\newcommand{\sklearn}{\textsc{scikit-learn}\xspace}

\newcommand{\expChem}{{Li$_{18}$Ta$_6$O$_{24}$}\xspace}
\newcommand{\expChemSG}{{$P2/c$}\xspace}

\newcommand{\LRtopOne}{\ensuremath{44.5}\xspace}
\newcommand{\CNNtopOne}{\ensuremath{70.0}\xspace}

\newcommand{\CNNtopN}{\ensuremath{91.9}\xspace}
\DeclarePairedDelimiter{\norm}{\lVert}{\rVert}
\newcommand{\dbNum}{\ensuremath{101, 802}\xspace}

\makeatletter
\newcommand{\floatcaption}{%
    \ifx \@captype \@undefined \@latex@error {\noexpand \caption outside float}\@ehd \expandafter \@gobble \else \refstepcounter \@captype \expandafter \@firstofone \fi {\@dblarg {\@caption \@captype }}%
}%
\makeatother

\journalcode{A}              

\begin{document}                  
    
    

\title{Using a machine learning approach to determine the space group of a structure from the atomic pair distribution function (PDF)}
\shorttitle{classifyPDF}

\author[a]{Chia-Hao}{Liu}
\author[a]{Yunzhe}{Tao}
\author[b]{Daniel}{Hsu}
\author[a]{Qiang}{Du}
\cauthor[a,c]{Simon J. L.}{Billinge}{sb2896@columbia.edu}

\aff[a]{Department of Applied Physics and Applied Mathematics, Columbia University, \city{New York}, New York, 10027, \country{USA}}
\aff[b]{Department of Computer Science, Columbia University, \city{New York}, New York, 10027, \country{USA}}
\aff[c]{Condensed Matter Physics and Materials Science Department, Brookhaven National Laboratory, \city{Upton}, New York, 11973, \country{USA}}

\maketitle
%

\begin{abstract}
We present a method for predicting the space group of a structure
given a calculated or measured atomic pair distribution function
(PDF) from that structure.
The method utilizes machine learning models trained on more than 100,000 PDFs calculated from structures in the 45 most heavily represented space groups.
In particular, we present a convolutional neural network (CNN) model which yields a promising result that it correctly identifies the space group among the top-6 estimates 91.9~\% of the time.
The CNN model also successfully identifies space groups on 12 out of 15 experimental PDFs.
We discuss interesting aspects of the failed estimates, which indicate that the CNN is failing in similar ways as conventional indexing algorithms applied to conventional powder diffraction data.
This preliminary success of the CNN model shows the possibility of model-independent assessment of PDF data on a wide class of materials.
    %
    
    \vspace{3em}
\end{abstract}

\section{Introduction}
Crystallography is used to determine crystal structures from diffraction patterns~\cite{giacovazzoDirectPhasingCrystallography1999}, including patterns from powdered samples~\cite{pecha;b;fopdascom05}.
The analysis of single crystal diffraction is the most direct approach for solving crystal structures.
However, powder diffraction becomes the best option when single crystals with desirable size and quality are not available.

A crystallographic structure solution makes heavy use of symmetry information to succeed.
The first step is to determine the unit cell and space group of the underlying structure.  Information about this is contained in the positions (and characteristic absences) of Bragg peaks in the diffraction pattern.
This process of determining the unit cell and space group of the structure is know as ``indexing" the pattern~\cite{giacovazzoDirectPhasingCrystallography1999}.
Indexing is inherently challenging for powder diffraction due to the loss of explicit directional information in the pattern, which is the result of projecting the data from three-dimensions into a one-dimensional pattern~\cite{dewolffDeterminationUnitcellDimensions1957,mighellGeometricalAmbiguitiesIndexing1975}.
However, there are a number of different algorithms available that work well in different situations~\cite{visserFullyAutomaticProgram1969,coelhoIndexingPowderDiffraction2003b,boultifPowderPatternIndexing2004,altomareAdvancesPowderDiffraction2009}
Once the unit cell information is determined, an investigation on systematic absences of diffraction peaks is carried out to identify the space group.
Various methods in determining space group information, based on either statistical or brute-force searches, have been used~\cite{neumannXCellNovelIndexing2003,markvardsenExtSymProgramAid2008a,altomareEXPO2009StructureSolution2009,coelhoIndexingAlgorithmIndependent2017}.

The problem is even more difficult when the structural correlations only extend on nanometer length-scales as crystallography breaks down~\cite{billi;s07}.
In this case progress can be made using atomic pair distribution function (PDF) methods for structure refinements~\cite{proff;zk05,egami;b;utbp12,choiStructureMethylammoniumLead2014b,zobel;s15,keen;n15}.
PDFs may also be used for studying structures of bulk materials.

There has been some success in using PDF for structure solution~\cite{juhas;n06,billi;4or18,juhas;jac10,cliff;prl10}.
However, a major challenge for PDF structure solution is that, unlike powder diffraction case, a peak in the PDF simply indicates a characteristic distance existing in the structure but no overall information about the underlying unit cell~\cite{egami;b;utbp12}.
Therefore, the symmetry information can not be inferred by the traditional indexing protocols that are predicated on the crystallography.
To date, there has been no demonstration of space-group determination directly from a PDF pattern.
Being able to determine the symmetry information based on the PDF will lead to more possibilities of solving structures from a wider class of materials.

Recently, machine learning (ML) has emerged as a powerful tool in different fields, such as in image classification~\cite{krizhevskyImageNetClassificationDeep2012} and speech recognition~\cite{hintonDeepNeuralNetworks2012}.
Moreover, ML models even outperform a human in cases such as image classifications~\cite{heDelvingDeepRectifiers2015} and the game of Go~\cite{silverMasteringGameGo2017}.
ML provides an platform of exploring the predictive relationship between the input and output of a problem, given a considerable amount of data is supplied for a ML model to ``learn''.
We know that the symmetry information is present in the powder diffraction pattern, and that the PDF is simply a Fourier transform of that pattern.
We therefore reason that the symmetry information survives in the PDF though we do not know explicitly how it is encoded.
We can qualitatively deduce that a higher symmetry structure, such as cubic, will produce a lower density of PDF peaks than a lower symmetry structure such as tetragonal.
However, it is beyond us to identify the space group directly, given the PDF.
Here we attempt to see whether a ML algorithm can be trained to recognise the space group of the underlying structure, given a PDF as input.
We note a recent paper that describes an attempt to determine the space group from powder diffraction pattern~\cite{parkClassificationCrystalStructure2017}.
In this case a promising accuracy of 81~\% was obtained in determining space group on simulated data, but the convolutional neural network (CNN) model they used was not able to determine space group from experimental data selected in their work.

To prepare data for training a ML model, we compute PDFs from \dbNum structures from 45 space groups deposited in the Inorganic Crystal Structure Database (ICSD) \cite{belskyNewDevelopmentsInorganic2002a}.
The space groups chosen were the most heavily represented, accounting for more than 80\% of known inorganic compounds \cite{urusovFrequencyDistributionSelection2009}.

The first ML model we tried was logistic regression (LR), which is a rather simple ML model.
Although quite successful, we explored a more sophisticated ML model, a convolutional neural network (CNN).
The CNN model outperforms the LR model by 15~\%, reaching an accuracy of \CNNtopN~\% for obtaining the correct space group in the top-6 predicted results on the testing set.
In particular, the CNN showed a significant improvement over LR in classifying challenging cases such as lower symmetry cases.

The CNN model is also tested on experimental PDFs where the underlying structures are known but the data are subject to experimental noise and collected under various instrumental conditions.
High accuracy in determining space groups from experimental PDFs was also demonstrated.

\section{The PDF method}
The experimental PDF, denoted $G(r)$, is the truncated Fourier transform of the total scattering structure function, $F(Q)=Q[S(Q) - 1]$~\cite{farro;aca09},
\begin{align}
\label{eq:FTofSQtoGr}
G(r) = \frac{2}{\pi} \int_{\qmin}^{\qmax} F(Q)\sin(Qr) \dd Q,
\end{align}
where $Q$ is the magnitude of the scattering momentum.
The structure function, $S(Q)$, is extracted from the Bragg and diffuse components of the powder diffraction intensity.
For elastic scattering, $Q = 4 \pi \sin(\theta) / \lambda$, where $\lambda$ is the scattering wavelength and $2\theta$ is the scattering angle.
In practice, values of $\qmin$ and $\qmax$ are determined by the experimental setup and $\qmax$ is often reduced below the experimental maximum to eliminate noisy data from the PDF since the signal to noise ratio becomes unfavorable in the high-$Q$ region.
The value of $\qmax$ is also known to be a dominant factor for the termination ripples introduced in the truncated Fourier transform~\cite{peter;jac03}.

The PDF gives the scaled probability of finding two atoms in a material at distance $r$ apart and is related to the density of atom pairs in the
material~\cite{egami;b;utbp12}.
For a macroscopic scatterer, $G(r)$ can be calculated from a known structure model according to
\begin{align}
\label{eq:Grfromrhor}
& G(r) = 4 \pi r \left[ \rho(r) - \rho_{0} \right],\\
& \rho(r) = \frac{1}{4 \pi r^{2} N}
\sum_{i}\sum_{j \neq i}
\frac{b_{i}b_{j}}{\langle b \rangle ^{2}}
\delta (r - r_{ij}).
\end{align}
Here, $\rho_{0}$ is the atomic number density of the material and $\rho(r)$ is the atomic pair density, which is the mean weighted density of neighbor atoms at distance $r$ from an atom at the origin.
The sums in $\rho(r)$ run over all atoms in the sample, $b_{i}$ is the scattering factor of atom $i$, $\langle b \rangle$ is the average scattering factor and $r_{ij}$ is the distance between atoms $i$ and $j$.

\section{Machine Learning experiments}
Machine learning (ML) is centered around the idea of exploring the predictive but oftentimes implicit relationship between inputs and outputs of a problem.
By feeding considerable amount of input and output pairs (training set) to a learning algorithm, we hope to arrive at a prediction model which is a good approximation to the underlying relationship between the inputs and outputs.
If the exact form of the output is available, either discrete or continuous, before the training step,  the problem is categorized as ``supervised learning''  under the context of ML.
The space-group determination problem discussed in this paper also falls into the supervised learning category.
In the language of ML, the inputs are often denoted as ``features'' of the data and the outputs are usually called the ``labels''.
Both inputs and outputs could be a scalar or a vector.
After learning the prediction model is then tested against a set of input and output pairs which have not seen by the training algorithm (the so-called  testing set) in order to independently validate the performance of the prediction model.

\begin{figure}
    \centering
    \includegraphics[width=0.8\columnwidth]{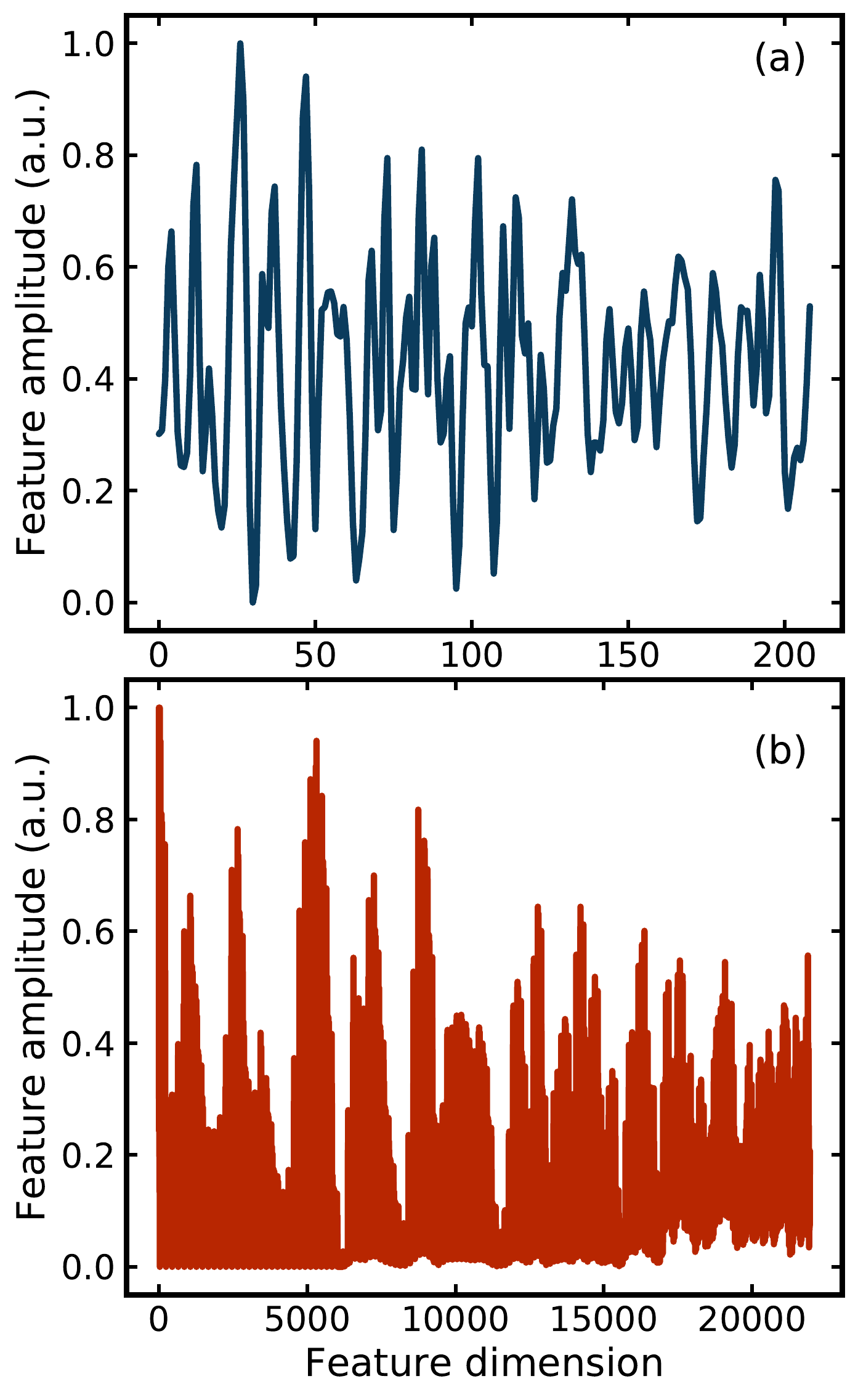}
    \caption{\label{fig:feature_eg}Example of (a) normalized PDF $\mathbf{X}$ and (b) its quadratic form $\mathbf{X}^2$ of compound \expChem (space group \expChemSG).}
\end{figure}

In the context of the space group determination problem, the input that we want to interrogate is PDF data.
We can select any feature or features from the data, for example, the feature we choose could be the PDF itself.
The label is the space group of the structure that gave rise to the PDF.
The database we will use to train our model is a pool of known structures.
Strictly, we choose all the known structures from 45 most heavily represented space groups in the ICSD, which accounts for 80~\% of known inorganic compounds~\cite{urusovFrequencyDistributionSelection2009}.
These were further pruned to remove duplicate (same composition \textit{and} same structure) entries.
The space groups considered and the number of unique structures in each space group are reproduced in Table~\ref{tab:sg_profile}.
\begin{table*}
    \centering
    \floatcaption{Space group and corresponding number of entries considered in this study.}
    \label{tab:sg_profile}
    \begin{tabular}{c|c}
        Space group (order) &  \# of entries \\
        \hline
        $P\bar{1} (2)$         &          4615 \\
        $P2_1 (4)$        &           581 \\
        $Cc (9)$          &           489 \\
        $P2_1/m (11)$     &          1247 \\
        $C2/m (12)$       &          3529 \\
        $P2/c (13)$       &           442 \\
        $P2_1/c (14)$     &          7392 \\
        $C2/c (15)$       &          3704 \\
        $P2_12_12_1 (19)$ &           701 \\
        $Pna2_1 (33)$     &           743 \\
        $Cmc2_1 (36)$     &           525 \\
        $Pmmm (47)$       &           646 \\
        $Pbam (55)$       &           745 \\
        $Pnnm (58)$       &           477 \\
        $Pbcn (60)$       &           478 \\
        $Pbca (61)$       &           853 \\
        $Pnma (62)$       &          6930 \\
        $Cmcm (63)$       &          2249 \\
        $Cmca (64)$       &           575 \\
        $Cmmm (65)$       &           513 \\
        $Immm (71)$       &           754 \\
        $I4/m (87)$       &           569 \\
        $I4_1/a (88)$     &           397 \\
        $I\bar{4}2d (122)$     &           373 \\
        $P4/mmm (123)$    &          1729 \\
        $P4/nmm (129)$    &          1376 \\
        $P4_2/mnm (136)$  &           870 \\
        $I4/mmm (139)$    &          4028 \\
        $I4/mcm (140)$    &          1026 \\
        $I4_1/amd (141)$  &           700 \\
        $R\bar{3} (148)$       &          1186 \\
        $R3m (160)$       &           482 \\
        $P\bar{3}m1 (164)$     &          1005 \\
        $R\bar{3}m (166)$      &          2810 \\
        $R\bar{3}c (167)$      &          1390 \\
        $P6_3/m (176)$    &          1289 \\
        $P6_3mc (186)$    &           849 \\
        $P6/mmm (191)$    &          3232 \\
        $P6_3/mmc (194)$  &          3971 \\
        $Pa\bar{3} (205)$      &           447 \\
        $F\bar{4}3m (216)$     &          2893 \\
        $Pm\bar{3}m (221)$     &          2933 \\
        $Fm\bar{3}m (225)$     &          4860 \\
        $Fd\bar{3}m (227)$     &          4382 \\
        $Ia\bar{3}d (230)$     &           455 \\
        \hline
        total & 101,802 \\
    \end{tabular}
\end{table*}
We then computed the PDF from each of \dbNum structures.
The parameters capturing finite $Q$-range and instrumental conditions, are reproduced in Table~\ref{tab:pdf_param}.
Those parameters are chosen such that they are close to the values that is practically attainable at most synchrotron facilities.
\begin{table*}
    \centering
    \floatcaption{Parameters used to calculate PDFs from atomic structures. All parameters follow the same definitions as in~\protect\cite{farro;jpcm07}.
        }
    \label{tab:pdf_param}
    {
    \begin{tabular}{c|c}
        Parameter & Value \\ \hline
        \rmin~(\AA) & 1.5\\
        \rmax~(\AA) & 30.0\\
        \qmax~(\AA$^{-1}$) & 0.5\\
        \qmin~(\AA$^{-1}$) & 23.0\\
        $r_{grid}$~(\AA) & $\frac{\pi}{\qmax}$\\
        ADP~(\AA$^{2}$) & 0.008\\
        $Q_{damp}$~(\AA$^{-1}$) & 0.04 \\
        $Q_{broad}$~(\AA$^{-1}$) & 0.01 \\
        \hline
    \end{tabular}
    }
\end{table*}
With the $r_{grid}$ and $r$-range reported in Table~\ref{tab:pdf_param}, each computed PDF is a $209\times1$ vector.
Depending on the atom types in the compounds, the amplitude of the PDF may vary drastically, which is inherently problematic for most ML algorithms~\cite{jamesIntroductionStatisticalLearning2013}
To avoid this problem, we determine a normalized PDF, $\mathbf{X}$ defined according to
\begin{align}
\label{eq:def_linear_feature}
\mathbf{X} = \frac{G(r)-\min(G(r))}{\max(G(r))-\min(G(r))},
\end{align}
where  $\min(\cdot)$ and $\max(\cdot)$ mean taking the minimum and maximum value of the target PDF function, $G(r)$, respectively.
Since $\min(\cdot)$ is always a negative number for the reduced PDF, $G(r)$, that we compute from the structure models, this definition results in the value of $\mathbf{X}$ always ranging between 0 and 1.
An example of $\mathbf{X}$ from \expChem (sapce group \expChemSG) is shown in Fig.~\ref{fig:feature_eg}(a).

For our learning experiments, we randomly select 80\% of the data entries from each space group as the training set and reserve the remaining 20\% of data entries as the testing set.

All learning experiments were carried out on one or multiple computation nodes of Habanero shared high performance cluster (HPC) at Columbia University.
Each computation node consists of 24 cores of CPUs (Intel® Xeon® Processor E5-2650 v4), 128GB memory and 2 GPUs (Nvidia K80 GPUs).

\subsection{Space Group Determination based on Logistic Regression}
We start our learning experiment with a rather simple model, logistic regression (LR).
In the setup of the LR model the probability of a given feature being classified as a particular space groups is parametrized by a ``logistic function''~\cite{hastieElementsStatisticalLearning2009}.
Forty-five space groups are considered in our study, therefore there are the same number of logistic functions, each with a set of parameters left to be determined.
Since the space group label is known for each data in the training set, the learning algorithm is then used to find an optimized set of parameters to each of the forty-five logistic functions such that the overall probability of determining the correct space group on all training data is maximized.
As a common practice, we also include ``regularization''~\cite{hastieElementsStatisticalLearning2009} to reduce overfitting in the trained model.
The regularization scheme chosen in our implementation is ``elastic net'' which is known for encouraging sparse selections on strongly correlated variables~\cite{zouRegularizationVariableSelection2005}.
Two hyperparameters $\alpha$ and $\Lambda$ are introduced under the context of our regularization scheme.
The explicit definition of these two parameters is presented in the Supporting Information section.
Our LR model is implemented through \sklearn~\cite{pedregosaScikitlearnMachineLearning2011}.
The optimum $\alpha, \Lambda$ for our LR model is determined by cross-validation~\cite{hastieElementsStatisticalLearning2009} in the training stage.

The best LR model with $\mathbf{X}$ as the input yields an accuracy of 20~\% at $(\alpha, \Lambda) = (10^{-5}, 0.75)$.
This result is better than a random guess from 45 space groups (2~\%) but is still far from satisfactory.
We reason that the symmetry information depends not on the absolute value of the PDF peak positions, which depend on specifics of the chemistry, but on their relative positions.
This information may be more apparent in an autocorrelation of the PDF with itself, which is a quadratic feature in ML language.
Our quadratic feature, $\mathbf{X}^2$, is defined as
\begin{align}
\label{eq:def_quadratic_feature}
\mathbf{X}^2 = \{\mathrm{X}_i\mathrm{X}_j~\vert~i, j = 1, 2, \dots d,~j > i \}
\end{align}
where $d$ is the dimension of $\mathbf{X}$ and $\mathbf{X}^2$ is a vector of dimension $\frac{d(d-1)}{2} \times 1$.
An example of the quadratic feature from \expChem (space group \expChemSG) is shown in Fig.~\ref{fig:feature_eg}(b).

\begin{figure}
    \centering
    \includegraphics[width=0.8\columnwidth]{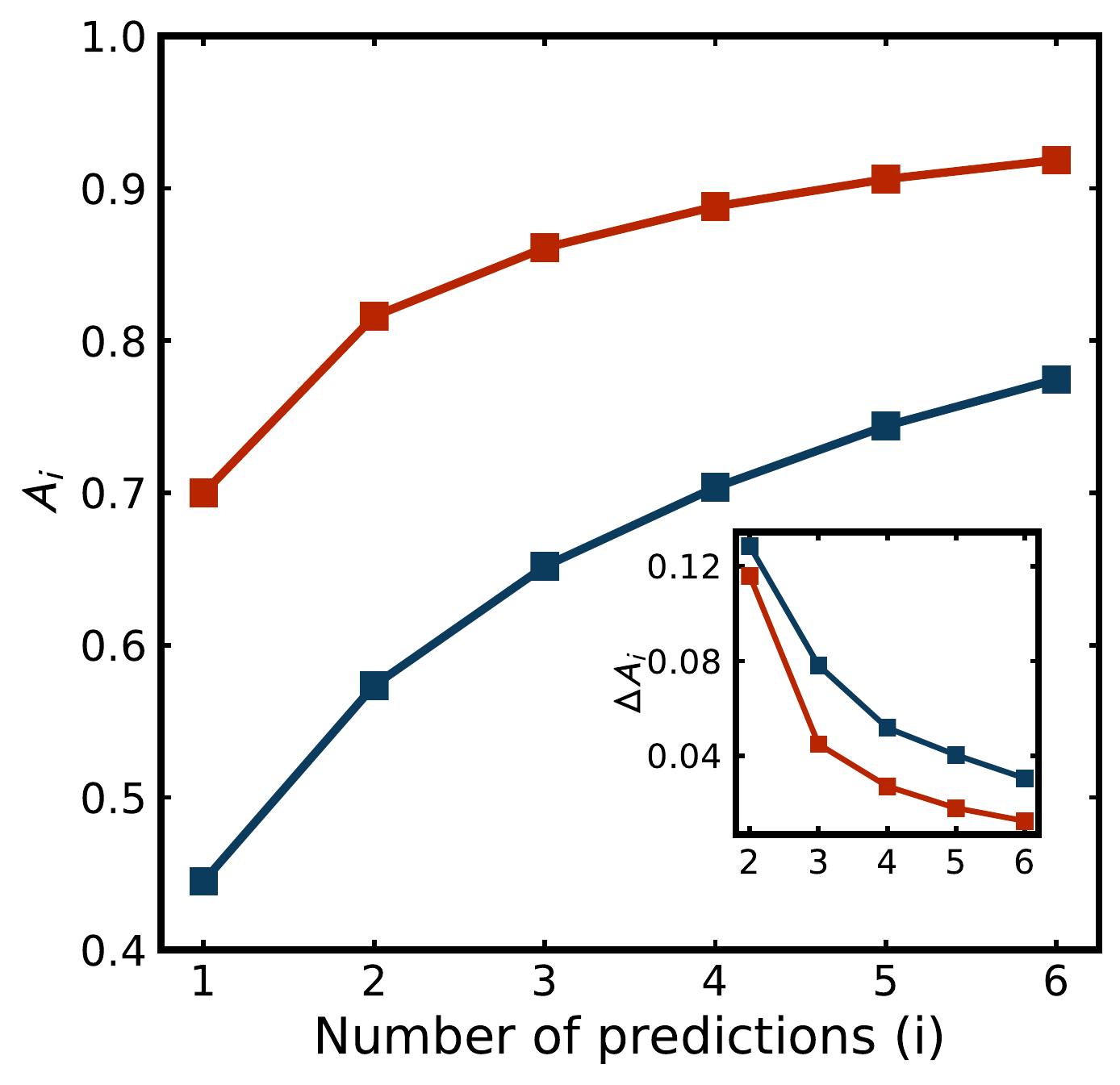}
    \caption{\label{fig:CF_top6_accu}Accuracy in determining space group when top-$i$ predictions are considered ($A_i$).
        Inset shows the first discrete differences ($\Delta A_i = A_i-A_{i-1}$) when $i$ predictions are considered.
        Blue represents the result of the logistic regression model with $\mathbf{X}^2$ and red is the result from the convolutional neural network model.}
\end{figure}

The best LR model with $\mathbf{X}^2$ as the input yields an accuracy
of \LRtopOne~\% at $(\alpha, \Lambda) = (10^{-5}, 1.0)$.
This is much better than for the linear feature, but still quite low.
However, the goal of space-group determination problem is to find the right space group, not necessarily to have it returned in the top position in a rank ordered list of suggestions.
We therefore define alternative accuracies that allow the correct space group to appear at any position in the top-3 ($A_3$) or top-6 ($A_6$) space groups returned by the model.
The values of $A_i$ $(i=1, 2, \dots 6)$ and their first discrete differences $\Delta A_i = A_i-A_In this case an accuracy of 81~\% was obtained in determining space group on simulated data, but the convolutional neural net (CNN) model they used was not able to determine space group from experimental data.{i-1}$ $(i=2, 3, \dots, 6)$ of our best LR model are shown in Fig.~\ref{fig:CF_top6_accu}.
We observed a more than 10~\% improvement in the alternative accuracy after considering top-2 predictions from the LR model ($\Delta A_2$) and the improvement ($\Delta A_i$) diminishes monotonically when more predictions are considered, as expected.
Top-6 is yielding a good accuracy (77~\%), and still a small enough number of space groups that could be tested manually in any structure determination.
%
%

\begin{figure}
    \includegraphics[width=0.8\columnwidth]{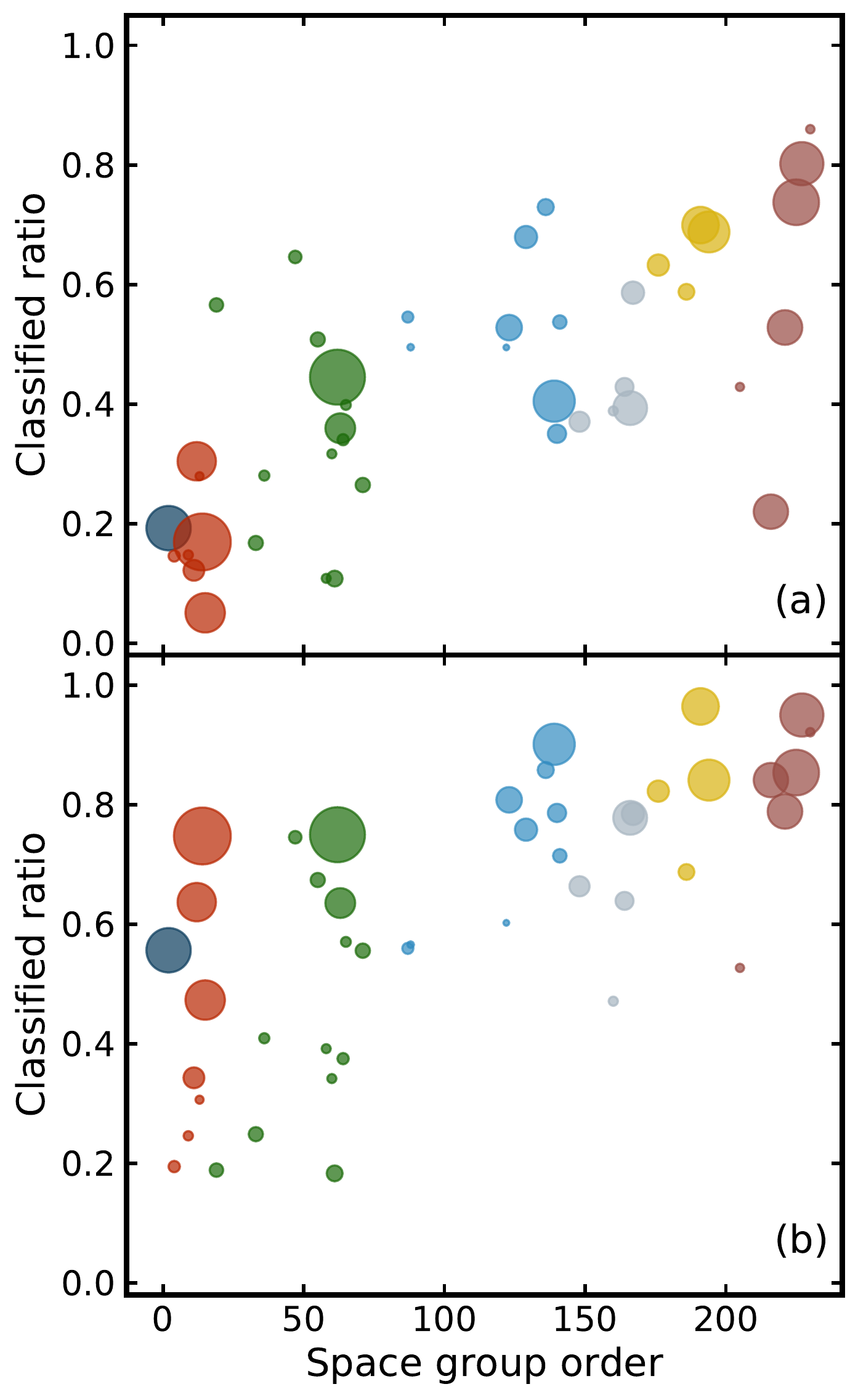}
    \caption{\label{fig:CF_classified_ratio}The ratio of correctly classified structures v.s. space group order from (a) logistic regression model (LR) with quadratic feature $\mathbf{X}^2$ and (b) convolutional neural network (CNN) model. Marker size reflects the relative frequency of space group in the training set. Markers are color coded with corresponding crystal systems (triclinic (dark blue), monoclinic (orange), orthorhombic (green), tetragonal (blue), trigonal (grey), hexagonal (yellow) and cubic (dark red).
    }
\end{figure}

The ratio of correctly classified structures vs. space group
order is shown Fig.~\ref{fig:CF_classified_ratio}(a).
Higher space group order means a more symmetric structure and we find, in general, the LR model yields a decent performance in predicting space groups from structures with high symmetry but it performs poorly on classifying low symmetry structures.

\subsection{Space group determination based on convolutional neural network (CNN)}
The result from the linear ML model (LR) is promising, prompting us to move to a more sophisticated deep learning model.
Deep learning models~\cite{lecunDeepLearning2015,goodfellowDeepLearning2016} have been successfully applied to various fields, ranging from computer vision~\cite{heIdentityMappingsDeep2016,krizhevskyImageNetClassificationDeep2012,radfordUnsupervisedRepresentationLearning2015}, natural language processing~\cite{bahdanauNeuralMachineTranslation2014,sutskeverSequenceSequenceLearning2014,kimConvolutionalNeuralNetworks2014} to material science~\cite{ramprasadMachineLearningMaterials2017,zilettiInsightfulClassificationCrystal2018}.
In particular, we sought to use a convolutional neural network (CNN)~\cite{lecunGradientbasedLearningApplied1998}.

The performance of a CNN depends on the overall architecture as well as the choice of hyperparameters such as the size of kernels, number of channels at each convolutional layer, the pooling size and the dimension of the fully-connected (FC) layer~\cite{goodfellowDeepLearning2016}.
However there is no well-established protocol for selecting these parameters, which is a largely trial and error effort for any given problem.
We build our CNN by trial-and-error, validating the performance on the testing data, which is just 20\% of the total data.

\begin{figure}
    \centering
    \includegraphics[width=1\columnwidth]{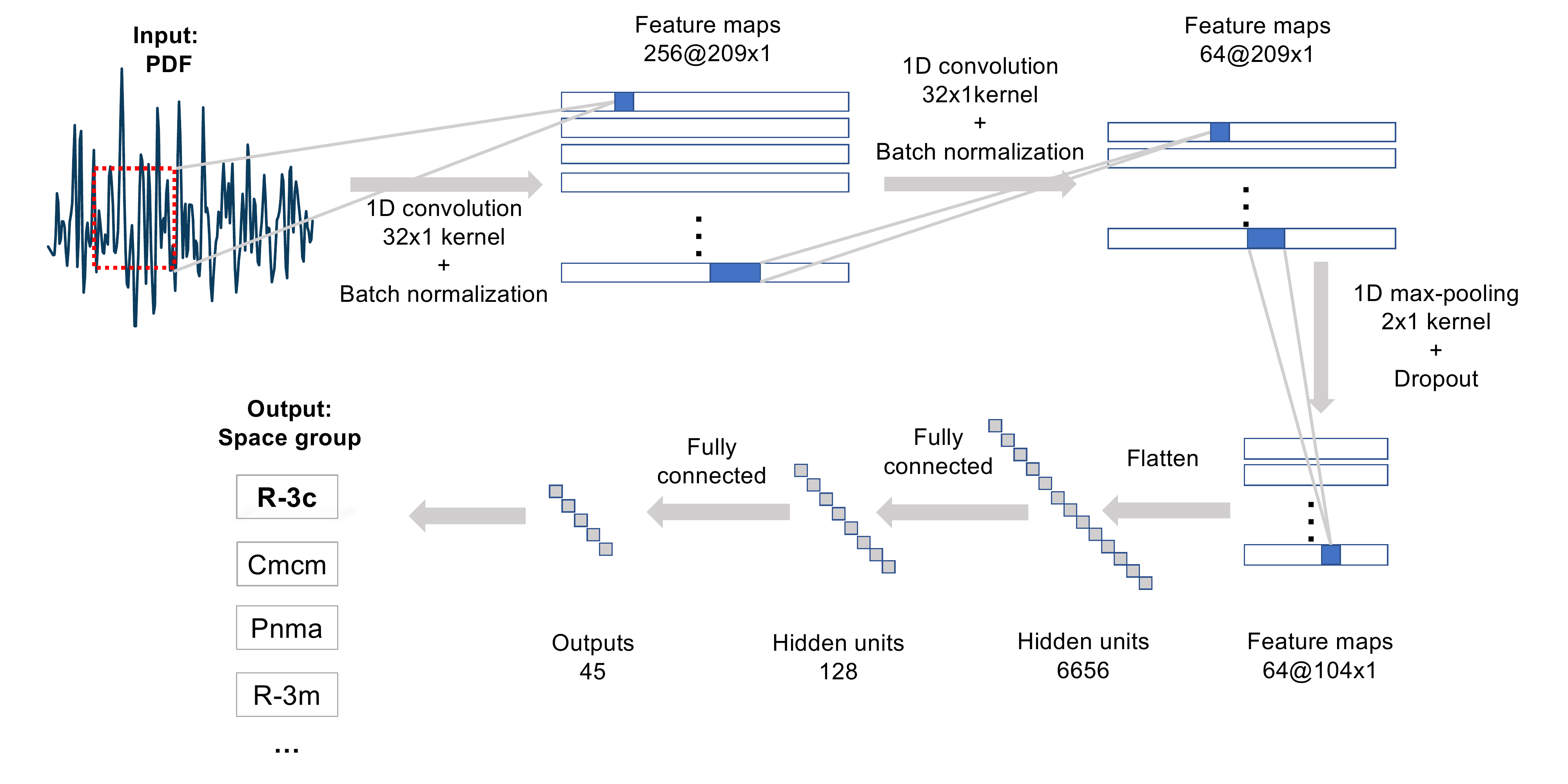}
    \label{fig:CNN_architecture}
    \caption{Schematic of our convolutional neural network (CNN) architecture.}
\end{figure}

The resulting CNN built for the space group determination problem is illustrated in Fig.~\ref{fig:CNN_architecture}.
The input PDF is a 1D signal sequence of dimension $209 \times 1 \times 1$.
We first apply a convolution layer of 256 channels with kernel size $32 \times 1$ to extract the first set of feature maps~\cite{lecunGradientbasedLearningApplied1998} of dimension $209 \times 1 \times 256$.
It has been shown that applying a nonlinear activation function to each output improves not only the ability for a model to learn complex decision rules but also the numerical stability during the optimization step~\cite{lecunDeepLearning2015}.
We chose rectified linear unit (ReLU)~\cite{dahlImprovingDeepNeural2013} as our activation function for the network.
After the first convolution layer, we apply 64-channel kernel of size $32 \times 1$ to the first feature map and generate the second set of feature maps of dimension $209 \times 1 \times 64$.
Similar to the first convolution layer, the second feature map is also activated by ReLU.
This is followed by a max-pooling layer~\cite{jarrettWhatBestMultistage2009} of size 2, which is applied to reduce overfitting.
After the subsampling process in the max-pooling layer, the output is of size $104 \times 1 \times 64$ and it is then flattened to size of $6556 \times 1$ before two fully-connected layers of size 128 and 45 are applied.
The first FC layer is used to further reduce the dimensionality of output from the max-pooling layer and it is activated with ReLu.
The second FC layer is activated with softmax function~\cite{goodfellowDeepLearning2016} to output the probability of the input PDF being one of the 45 space groups considered in our study.

Categorical cross entropy loss~\cite{bishopPatternRecognitionMachine2006} is used for training our model.
It is apparent from Table~\ref{tab:sg_profile} that the number of data entries in each space group are not evenly distributed, varying from 373 ($I\bar{4}2d$) to 7392 ($P2_1/c$) per space group.
We would like to avoid the possibility that we obtain a neural network that is biased towards space groups with more abundant data entries.
To mitigate the effect of the unbalanced data set, loss from each training sample is multiplied by a class weight~\cite{kingLogisticRegressionRare2001} which is the inverse of the ratio between the number of data entries from the same space group label as the training sample and the size of entire training set.
We then use Adaptive Moment Estimation (Adam)~\cite{kingmaAdamMethodStochastic2014} as the stochastic optimization method to train our model with a mini-batch size of~64.
During the training step, we follow the same protocol outlined in Ref.~\cite{heIdentityMappingsDeep2016} to perform the weight initialization~\cite{heDelvingDeepRectifiers2015} and batch normalization~\cite{ioffeBatchNormalizationAccelerating2015}.
A dropout strategy~\cite{srivastavaDropoutSimpleWay2014} is also applied in the pooling layer to reduce overfitting in our neural network.
The parameters in the CNN model are iteratively updated through the stochastic gradient descent method (Adam).

\begin{figure}
    \centering
    \includegraphics[width=1\columnwidth]{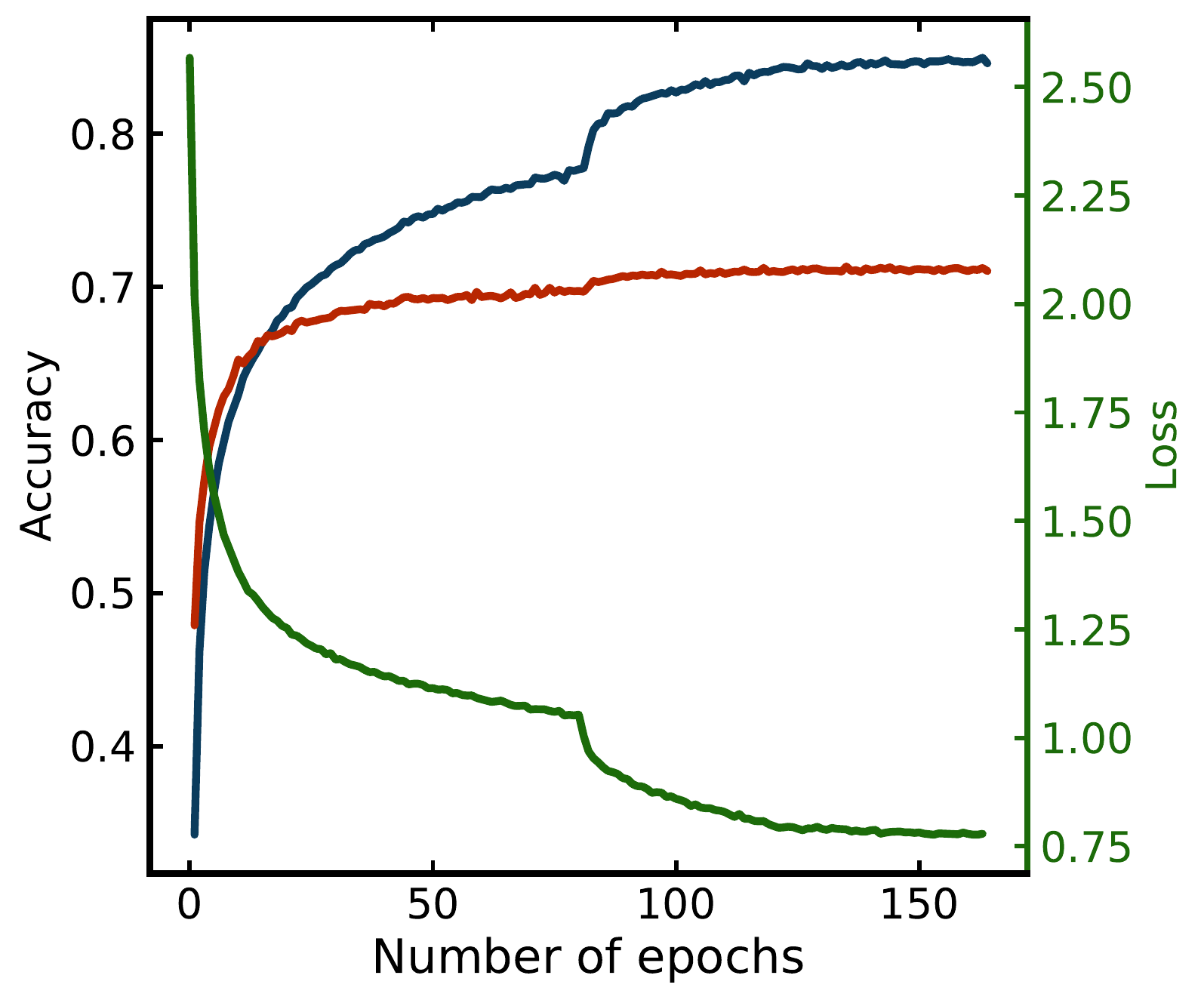}
    \caption{\label{fig:CNN_learning_curve}Accuracy of the CNN model on the training set (blue), the testing set (red) and the optimization loss against the testing set (green) with respect to number of epochs during the training step.}
\end{figure}

Learning rate is a parameter that affects how drastically the parameters are updated at each iteration.
A small learning rate is preferable when the parameters are close to some set of optimal values and vice versa.
Therefore, an appropriate schedule of learning rate is crucial for training a model.
Our training starts with a learning rate of 0.1, and the value is reduced by a factor of 10 at epochs~81 and then~122.
With the learning rate schedule described, the optimization loss against the testing set, along with the prediction accuracy on the training and testing sets, are plotted with respect to the number of epochs in Fig.~\ref{fig:CNN_learning_curve}.
Our training is terminated after~164 epochs when the training accuracy, testing accuracy and optimization loss all plateau, meaning no significant improvement to the model would be gained with further updates to the parameters.
Our CNN model is implemented by Keras~\cite{chollet2015keras} and trained on a single Nvidia Tesla K80 GPU.
Under the architecture and training protocol discussed above, our best CNN model yields an accuracy of \CNNtopOne\% from top-1 prediction and \CNNtopN\% from top-6 predictions, which outperforms the LR model by $15~\%$.
Similarly, from Fig.~\ref{fig:CF_top6_accu}, we observe a more than 10~\% improvement in the alternative accuracy after considering top-2 predictions ($\Delta A_2$) in the CNN model and the improvement ($\Delta A_i$) decreases monotonically, even in a more drastic trend than the case of LR model, when more predictions are considered.

\section{Discussion}
In general, it is fair to expect a ML model to achieve a higher accuracy on a space group with abundant training samples.
However, from Fig.~\ref{fig:CF_classified_ratio}, it is clear that the LR model even fails to identify well represented space groups across all space group orders.
On the other hand, a positive correlation between the size of training data and the classification ratioIn this case an accuracy of 81~\% was obtained in determining space group on simulated data, but the convolutional neural net (CNN) model they used was not able to determine space group from experimental data. is observed in the CNN model.
Furthermore, except for space group Ia-3d which is the most symmetric space group, the classification ratios on the rarely seen groups are lower than the well represented groups in our CNN model.
However, the main result is that the CNN performs significantly better than the LR model for all space groups, especially on structures with lower symmetry.
There is an overall trend towards increase in the prediction ability as the symmetry increases, and there are outliers, but there seems to be a trend that the model is better at predicting space groups for more highly populated space groups.

\begin{figure}
    \centering
    \includegraphics[width=1\textwidth]{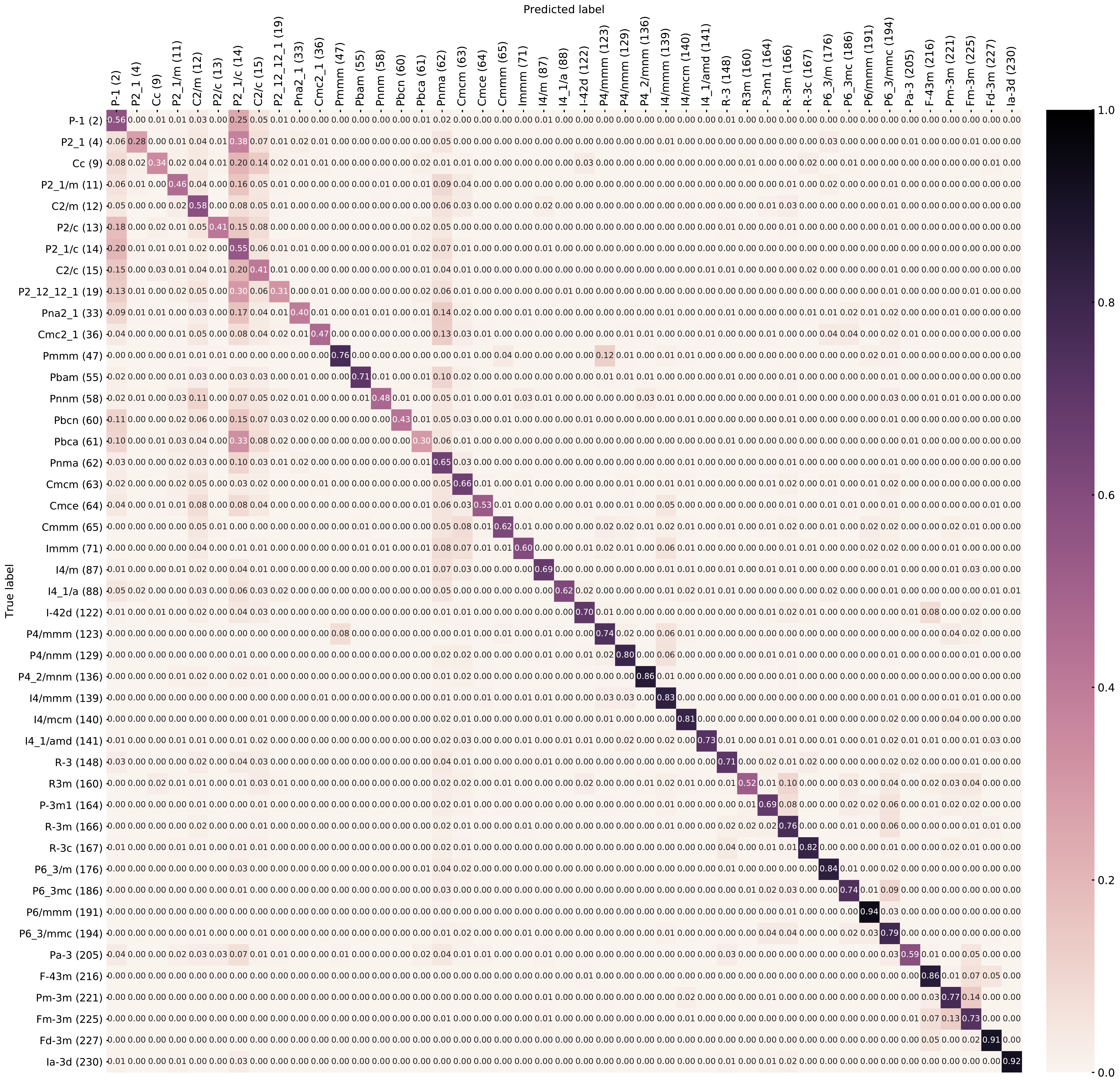}
    \caption{\label{fig:CNN_confusion_mat}The confusion matrix of our CNN model. The row labels indicate the correct space group and the column labels the space group returned by the model. An ideal model would result in a confusion matrix with all diagonal values being 1 and all off-diagonal values being zero.  The numbers in parentheses are the space-group number.
    }
\end{figure}

The confusion matrix~\cite{stehmanSelectingInterpretingMeasures1997} is a common tool to assess the performance of a ML model.
The confusion matrix, $\mathbf{M}$, is an $N$ by $N$ matrix, where $N$ is the number of labels in the dataset.
The rows of $\mathbf{M}$ identify the true label (correct answer) and the columns of $\mathbf{M}$ mean the label predicted by the classifier.
The numbers in the matrix are the proportion of results in each category.
For example, the diagonal elements indicate the proportion of outcomes where the correct label was predicted in each case, and
the matrix element in the $Fd\bar{3}m$ row and the $F\bar{4}3m$ column (value 0.05) is the proportion of PDFs from an $Fd\bar{3}m$ space group structure that were incorrectly classified as being in space group $F\bar{4}3m$.
For an ideal prediction model, the diagonal elements of the confusion matrix should be $1.0$ and all off-diagonal elements would be zero.
The confusion matrix from our CNN classifier is shown Fig.~\ref{fig:CNN_confusion_mat}.

We observe ``teardrop'' patterns on the columns of $P\bar{1}$, $P2_1/c$ and $Pnma$, meaning the CNN model tends to incorrectly assign a wide range of space groups into these groups.
On the surface, this behavior is worrying but the confusions actually correspond to the real group-subgroup relation which has been known and tabulated in the literature~\cite{ascherKorrekturenAngabenUntergruppen1969,boyleKlassengleichenSupergroupSubgroup1972,hahnInternationalTablesCrystallography2002}
For the case of $P\bar{1}$, the major confusion groups ($P2_1/c$, $C2/c$ and $P2/c$) are in fact minimal non-isomorphic supergroups of $P\bar{1}$.
Moreover, $P2_12_12_1$ shares the same subgroup ($P2_1$) with $P2_1/c$ and $Pbca$ is a supergroup of $P2_12_12_1$ while $Pbcn$ is a supergroup of $P2_1/c$.
Similar reasoning can be applied to the case of $P2_1/c$ and $Pnma$ as well.
The statistical model appears to be picking up some real underlying mathematical relationships.

We also investigate the cases with low classification accuracy (low value in diagonal elements) from the CNN model.
$P2_1$ is the group with the lowest accuracy (28~\%) among all labels
The similar group-subgroup reasoning also holds for this case as well.
$P2_1/c$ (38~\% error rate) is, again, a supergroup of $P2_1$ and $C2/c$ (7~\% error rate) is a supergroup of $P2_1/c$.
The same reasoning holds for other confusion cases and we will not explicitly iterate through it here, but this suggests that these closely group/sub-group related space groups should also be considered whenever the classifier returns another one in the series.
It is possible to train a different CNN model which focuses on disambiguating space groups that are closely related by the group/sub-group relationship.
However, we did not implement this kind of hierarchical model in our study.

\section{Space Group Determination on Experimental PDFs}
The CNN model is used to determine the space group of 15 experimental PDFs and the results are reported in Table~\ref{tab:expData_test}.
For each experimental PDF, structures are known from previous studies which are also referenced in the table.
Both crystalline (C) or nanocrystalline (NC) samples with a wide range of structural symmetries are covered in this set of experimental PDFs.
It is worth noting that the sizes of the NC samples chosen are roughly equal to or larger than 10~nm, at which size in our measurements the PDF signal from the NC material falls off roughly at the same rate as that from crystalline PDFs in the training set due.
Every experimental PDF is subject to experimental noise and collected under various instrumental conditions that result in aberrations to the PDF that are not identical to parameter values used to generate our training set (Table~\ref{tab:pdf_param}).
It is therefore expected that the CNN classifier will work less well than on the testing set.
From Table~\ref{tab:expData_test}, it is clear that the CNN model yields an overall satisfactory result in determining space groups from experimental data with the space group from 12 out of 15 test cases properly identified in the top-6 predictions.

Here we comment on the performance of the CNN.
In the cases of IrTe$_2$ at 10~K, the material has been reported in the literature in both $C2/m$ and $P\bar{1}$ space groups, and it is not clear which is correct.
The CNN returned both space groups in the top six.
These space groups are known to be difficult to differentiate~\cite{matsumotoResistanceSusceptibilityAnomalies1999c,toriyamaSwitchingConductingPlanes2014a}.
Furthermore, for data from the same sample at room temperature, the CNN model identifies not only the correct space group ($P\bar{3}m1$), but also the space groups that the structure will occupy below the low-temperature symmetry lowering transition ($C2/m$, $P\bar{1}$).
For the case of BaTiO$_3$ nanoparticles, the CNN model identifies two space groups that are considered in the literature to yield rather equivalent structures ($R3m$, $P4/mmm$)~\cite{pageProbingLocalDipoles2010d,polkingFerroelectricOrderIndividual2012c}.
It is encouraging that the CNN appears to be getting the physics right in these cases.

Investigating the failing cases from the CNN model (entries with a dagger in Table~\ref{tab:expData_test}) also reveals insights about the decision rules learned by the model.
Sr$_2$IrO$_4$, was firstly identified as a perovskite structure with space group $I4/mmm$~\cite{randallPreparationStrontiumIridiumOxide1957}, but later work pointed out that a lower symmetry group $I4_1/acd$ is more appropriate due to correlated rotations of the corner-shared IrO$_6$ octahedra about the $c$-axis~\cite{huangNeutronPowderDiffraction1994,shimuraStructureMagneticProperties1995}.
There is a long-wavelength modulation of the rotations along the $c$-axis resulting a supercell with a five-times expansion along that direction ($a=5.496$~\AA, $c=25.793$~\AA).
The PDF will not be sensitive to such a long-wavelength superlattice modulation which may explain why the model does not identify a space group close to the $I4_1/acd$ space group, reflecting additional symmetry breaking due to the supermodulation.
It is not completely clear what the space group would be for the rotated octahedra without the supermodulation, so we are not sure if this space group is among the top-6 that the model found.

Somewhat surprisingly the CNN fails to find the right space group for wurtzite CdSe, which is a very simple structure, but rather finds space groups with low symmetries.
One possible reason is that we know there is a high degree of stacking faulting in the bulk CdSe sample that was measured.
This was best modelled as a phase mixture of wurtzite (space group $P6_3mc$) and zinc-blende (space group $F\bar{4}3m$)~\cite{masad;prb07}.
The prediction of low symmetry groups might reflect the fact the underlying structure can not be described with a single space group.

\begin{table*}
    \centering
    \floatcaption{\label{tab:expData_test}Top-6 space group predictions results from the CNN model with experimental PDFs.
        Bold-faced prediction is the most probable space group from existing literatures listed in the Refs. column.
        More than one predictions are highlighted when these space groups are regarded as highly similar in literatures.
        Details about these cases will be discussed in the text.
        The Note column specifies if the PDF is from a crystalline (C) or nanocrystalline (NC) material.
        The experimental data were collected under various instrumental conditions which are  not identical to the training set and experimental data were measured at the room temperature, unless otherwise specified.
        Dagger is used to label the data that the CNN model fails to predict the correct space group.
    }
    \vspace{1em}
    \resizebox{1\columnwidth}{!}{
    \begin{threeparttable}
        \begin{tabular}{c|cccccc|c|c}
            Sample      & 1st                & 2nd             & 3rd       & 4th           & 5th           & 6th           & Refs.  &Note          \\
            \hline
            Ni          & $\boldsymbol{Fm\bar{3}m}$     & $Pm\bar{3}m$           & $Fd\bar{3}m$     & $F\bar{4}3m$        & $P4/mmm$        & $P6_3/mmc$     & \cite{owenLXVIXrayMeasurement1936} & C               \\
            Fe$_3$O$_4$       & $\boldsymbol{Fd\bar{3}m}$     & $I4_1/amd$       & $R\bar{3}m$      & $Fm\bar{3}m$         & $F\bar{4}3m$         & $P6_3/mmc$     &  \cite{fleetStructureMagnetite1981} & C             \\
            CeO$_2$        & $\boldsymbol{Fm\bar{3}m}$     & $Fd\bar{3}m$           & $Pm\bar{3}m$     & $F\bar{4}3m$         & $Pa\bar{3}$          & $P4/mmm$        & \cite{yashimaPositionalDisorderOxygen2004} & C               \\
            Sr$_2$IrO$_4^\dagger$     & $Fm\bar{3}m$              & $P6/mmm$          & $P6_3/mmc$ & $Pm\bar{3}m$         & $Fd\bar{3}m$         & $R\bar{3}m$          &\cite{huangNeutronPowderDiffraction1994,shimuraStructureMagneticProperties1995} & C   \\
            CuIr$_2$S$_4$     & $\boldsymbol{Fd\bar{3}m}$     & $Fm\bar{3}m$           & $F\bar{4}3m$     & $R\bar{3}m$          & $Pm\bar{3}m$         & $R3m$           & \cite{furubayashiStructuralMagneticStudies1994b} & C  \\
            CdSe$^\dagger$  & $P2_1/c$            & $P\bar{1}$             & $C2/c$      & $Pnma$          & $Pna2_1$       & $P2_12_12_1$ & \cite{masad;prb07}    & C  \\
            IrTe$_2$  & $C2/m$               & $\boldsymbol{P\bar{3}m1}$  & $P2_1/c$   & $P\bar{1}$           & $P2_1/m$       & $C2/c$          &  \cite{matsumotoResistanceSusceptibilityAnomalies1999c} & C              \\
            IrTe$_2$@10K   & $\boldsymbol{C2/m}$               & $P6_3/mmc$       & $P6/mmm$    & $P4/mmm$        & $\boldsymbol{P\bar{1}}$  & $P2_1/c$       & \cite{matsumotoResistanceSusceptibilityAnomalies1999c,toriyamaSwitchingConductingPlanes2014a}  & C              \\
            Ti$_4$O$_7$       & $\boldsymbol{P\bar{1}}$       & $C2/c$            & $P2_1/c$   & $C2/m$          & $Pnnm$          & $P4_2/mnm$     & \cite{marezioCrystalStructureTi4O71971} & C               \\
            MAPbI$_3$@130K & $P\bar{1}$                & $P2_1/c$         & $C2/c$      & $P2_12_12_1$ & $\boldsymbol{Pnma}$ & $Pna2_1$       & \cite{swainsonPhaseTransitionsPerovskite2003} & C               \\
            MoSe$_2$       & $\boldsymbol{P6_3/mmc}$ & $R3m$             & $R\bar{3}m$      & $P6_3mc$       & $P4/mmm$        & $Fd\bar{3}m$         &  \cite{jamesCrystalStructureMoSe21963} & C\\
            \hline
            TiO$_2$(anatase)     & $\boldsymbol{I4_1/amd}$ & $C2/m$            & $P2_1/m$   & $C2/c$          & $P\bar{1}$           & $P2_1/c$       &\cite{hornRefinementStructureAnatase1972a}    &  NC         \\
            TiO$_2$(rutile)      & $\boldsymbol{P4_2/mnm}$ & $C2/m$            & $P2_1/c$   & $P\bar{1}$           & $P2_1/m$       & $Pnma$          & \cite{baurRutiletypeCompoundsIV1971a}  & NC              \\
            Si$^\dagger$       & $P6_3mc$            & $I\bar{4}2d$           & $R3m$       & $C2/c$          & $P\bar{1}$           & $Pbca$          & \cite{rohani;acsn18} & NC \\
            BaTiO$_3$     & $\boldsymbol{R3m}$       & $\boldsymbol{P4/mmm}$ & $C2/m$      & $P6_3/mmc$     & $Pnma$          & $Cmcm$          & \cite{pageProbingLocalDipoles2010d,polkingFerroelectricOrderIndividual2012c}] & NC
        \end{tabular}
    \end{threeparttable}
    }
\end{table*}

\section{Conclusion}
We demonstrate an application of machine learning (ML) to determine the space group directly from an atomic pair distribution function (PDF).
We also present a convolutional neural network (CNN) model which yields a promising accuracy (\CNNtopN~\%) from the top-6 predictions when it is evaluated against the testing data.
Interestingly, the trained CNN model appears to capture decision rules that agree with the mathematical (group-subgroup) relationships between space groups.
The trained CNN model is tested against 15 experimental PDFs, including crystalline and nanocrystalline samples.
Space groups from 12 of these experimental data were successfully found in the top-6 predictions by the CNN model.
This shows great promise for preliminary, model-independent assessment of PDF data from well ordered crystalline or nanocrystalline materials.

\section{Acknowledgements}
X-ray PDF measurements were conducted on beamline 28-ID-1 (PDF) and 28-ID-2 (XPD) of the National Synchrotron Light Source II, a U.S. Department of Energy (DOE) Office of Science User Facility operated for the DOE Office of Science by Brookhaven National Laboratory under Contract No. DE-SC0012704.

\bibliographystyle{iucr}
\bibliography{liu_classifypdf_iucr}


\setcounter{figure}{0}
\setcounter{equation}{0}
\setcounter{table}{0}
\makeatletter
\renewcommand{\fnum@figure}{Fig.~S\thefigure}
\renewcommand{\theequation}{S\arabic{equation}}
\renewcommand{\thetable}{S\arabic{table}}
\makeatother

\appendix
\section{Logistic Regression and Elastic Net Regularizations}

Consider a dataset with total $M$ structures and $K$ distinct space-group labels.
Each structure has a space group and we denote the space group of $m$-th structure as $k_m$ where $k_m \in \{1, 2, \dots K\}$, our complete set of space groups.
In the setup of LR model, the probability of a feature $x_m$ of dimension $d$, which is a computable from $m$-th structure, belongs to a specific space group $k_m$ is parametrized as
\begin{align}
\label{eq:logistic_prob}
\mathrm{Pr}(k_m|x_m, \beta^{k_m}) = \frac{\exp\left(\beta_0^{k_m} + \sum \limits_{i=1}^{d} \beta_i^{k_m} x_{m,i}\right)}{1+\exp\left(\beta_0^{k_m} + \sum \limits_{i=1}^{d} \beta_i^{k_m} x_{m,i}\right)},
\end{align}
where $\beta^{k_m} =\{\beta_0^{k_m}, \beta_1^{k_m}, \dots, \beta_d^{k_m} \}$ is a set of parameters to be determined. 
The index $k_m$ runs from 1 to 45 which corresponds total number of space groups considered in our study.
Since the space group $k$ and feature $x$ are both known for the training data, the learning algorithm is then used to find a optimized set of $\beta = \{\beta^{k_m}: k_m = 1, 2, \dots, K\}$ which maximizes the overall probability in determining correct space group $\mathrm{Pr}(k_m|x_m,  \beta^{k_m})$ on all $M$ training data.

For each of the $M$ structures, there will be a binary result for classification; Either the space group label is correctly classified or not.
This process can be regarded as $M$ independent Bernoulli trials. The probability function for a single Bernoulli trial is expressed as
\begin{align}
f(k_m | x_m,
\boldsymbol{\beta}^{k_m}) = &\left[\mathrm{Pr}(k_m | x_m, \boldsymbol{\beta}^{k_m})\right]^{\gamma_m}\\ \nonumber \nonumber &\left[1-\mathrm{Pr}(k_m | x_m, \boldsymbol{\beta}^{k_m})\right]^{1-\gamma_m},
\end{align}
where $\gamma$ is an indicator. $\gamma_m = 1$ if the space-group label $k_m$ is correctly predicted and $\gamma_m = 0$ if the prediction is wrong.
Since each classification are independent, the joint probability function for $M$ classifications on the space-group label, $f_M(\boldsymbol{K} | x, \boldsymbol{\beta})$, is written as
\begin{align}
\label{eq:bernoulli_trails_prob}
f_M(\mathbf{K} | \mathbf{x}, \boldsymbol{\beta}) = \prod_{m=1}^{M} f(k_m | x_m, \boldsymbol{\beta}^{k_m}),
\end{align}
where $\mathbf{K} = \{k_m\}$ and $\mathbf{x}=\{x_m\}$.
Furthermore, since both the label and features are known in the training set, Eq.~\ref{eq:bernoulli_trails_prob} is just a function of $\beta$,
\begin{align}
\label{eq:likelihood}
    L(\beta) = f_M(\mathbf{K} | \mathbf{x}, \boldsymbol{\beta})
\end{align}
Logarithm is a monotonic transformation.
Taking logarithm of Eq.~\ref{eq:likelihood} does not change the original behavior of the  function and it improves the numerical stability as the product of probabilities is turned into sum of logarithm of probabilities and extreme values from the product can still be computed numerically.
We therefore arrive the ``log-likelihood'' function
\begin{align}
\label{eq:log_likelihood}
    l(\beta) = \log (L(\beta))
\end{align}
It is common to include ``regularization''~\cite{hastieElementsStatisticalLearning2009} for reducing overfitting in the model.
The regularization scheme chosen in our implementation is ``elastic net'' which is known for encouraging sparse selections on strongly correlated variables~\cite{zouRegularizationVariableSelection2005}.
The explicit definitions of the log-likelihood function with elastic regularization is written as
\begin{align}
\label{eq:logistic_full_loss}
l_t(\beta) = l(\beta) + \alpha \Bigl(\Lambda \norm{\beta}_1 + (1-\Lambda) \norm{\beta}_2^2 \Bigr),
\end{align}
where $\norm{\cdot}$ and  $\norm{\cdot}_2^2$ stands for L1 and L2 norm~\cite{hornMatrixAnalysisSecond2012} respectively.
Two hyperparameters $\alpha$ and $\Lambda$ are introduced under this regularization scheme.
$\alpha$ is a hyperparameter that  determines the overall ``strength'' of the regularization and $\Lambda$ governs the relative ratio between L1 and L2 regularization~\cite{zouRegularizationVariableSelection2005}.
Detailed steps on optimizing Eq.~\ref{eq:logistic_full_loss} is beyond the scope of this paper, but they are available in most of standard ML reviews~\cite{hastieElementsStatisticalLearning2009,bishopPatternRecognitionMachine2006}.

\section{Robustness of the CNN model}
The classification accuracies from CNN models with different sets of hyperparameters, such as number of filters, kernel size and pooling size, are reproduced in Table~\ref{tab:cnn_hyperparam}.
\begin{table*}
    \centering
    \floatcaption{Accuracies of CNN model with different sets of hyper parameters. Accuracy is abbreviated as accu. in the table. The last row specifies the optimum set of hyperparmeters for our final CNN model.}
    \label{tab:cnn_hyperparam}
    \vspace{1em}
    \begin{tabular}{c|c|c|c|cc}
        \# filters & kernel size & \# hidden units & \# ensembles & Top-1 accu. (\%) & Top-6 accu. (\%) \\
        \hline
        128, 32    & 24          & 128             & 2            & 64.1        & 90.7        \\
        256, 64    & 24          & 128             & 2            & 68.6        & 91.6        \\
        64, 64     & 24          & 128             & 2            & 67.4        & 91.1        \\
        \hline
        128, 64    & 32          & 128             & 2            & 69.0        & 91.7        \\
        128, 64    & 16          & 128             & 2            & 66.6        & 91.3        \\
        \hline
        128, 64    & 24          & 256             & 2            & 69.2        & 91.6        \\
        128, 64    & 24          & 64              & 2            & 66.4        & 91.2        \\
        \hline
        128, 64    & 24          & 128             & 1            & 65.7       & 91.1       \\
        128, 64    & 24          & 128             & 3            & 68.2        & 91.6        \\
        \hline
        \textbf{256, 64}    & \textbf{32}          & \textbf{128}             & \textbf{3}            & \textbf{70.0}        & \textbf{91.9}
     \end{tabular}
\end{table*}
        The classification accuracy only vary modestly across different sets of hyperparameters and this implies the robustness of our CNN architecture.
        We determined the desired architecture of our CNN model based on the classification accuracy on the testing set and the learning curves (loss, training accuracy and testing accuracy) reported in Fig.~\ref{fig:CNN_learning_curve}.
\end{document}